# The stochastic gravitational-wave background exists permanently and demonstrates time-domain asymmetry

Alexander V. Kramarenko[1], Andrey V. Kramarenko[2,3,*]

[1]TREDEX Company Ltd., PO box 1515, Kharkiv, Ukraine 61001; tredexcompany37@gmail.com

[2]National Technical University "KhPI", 2 Kyrpychova str., Kharkiv, Ukraine 61002; andrii.kramarenko@khpi.edu.ua

[3] V. N. Karazin Kharkiv National University, 4 Svobody Sq., Kharkiv, Ukraine 61022; kramarenko@karazin.ua

## Abstract

Analyzing the records of Advanced LIGO and Virgo gravitational observatories, we found a permanent time-domain asymmetry inherent only to the signals of their gravitational detectors. Experiments with different periodic signals, Gaussian and non-Gaussian noises, made it possible to conclude that the noise of gravitational detectors is an unusual mixture of signals. We also developed a specialized Pearson correlation analyzer to detect the gravitational-wave (GW) events. It turned out that the LIGO and Virgo detectors' output signals contain a significant ($-6$ dB) component demonstrating the properties of records of confirmed resolved GW events. It allows us to argue that the gravitational background is largely due to the processes of stellar masses merging. Since the specific signal is registered by the detectors continuously, we can consider the sub-kilohertz band gravitational background field as discovered. Our analysis method also allows us to estimate the contribution of the gravitational background component to the total signal energy. With its help, it will be possible not only to provide the radio-frequency estimation of the magnitude of gravitational disturbances but also to obtain the GW background sky map.

## Introduction

By now, the gravitational-wave events detected by the Advanced LIGO[1] and Virgo[2] observatories are the pulse signals having combined frequency and amplitude modulation, smooth rise and sharp roll-off[3]. The linear frequency modulated (LFM) or "chirp" radar signals[4] have a similar time-domain asymmetry in frequency only while the pulse amplitude does not change significantly.

The optimal detector of a signal with such a specific waveform would be convolution with its time-reversed reference pattern[5], since a convolution with a direct pattern yields a much smaller response that is not concentrated in time due to the asymmetry of the pulses in the time domain.

As far as we know, the question about the contribution of gravitational waves due to the merging of rotating masses to the stochastic background gravitational-wave noise is currently topical, widely discussed in a number of articles[6–11], and far from being finally solved. We should separately mention the article[7], the authors of which proposed a method of detecting stellar mass merger events based on small differences between the observed noise and the Gaussian noise. But the Gaussian criteria in no way forbid noise to be symmetric or asymmetric in time.

Let us take a sufficiently large realization of ideal Gaussian white noise and divide it into a set of equal fragments calculating the median frequency of the spectrum of each fragment. Then assemble a new realization from the fragments so that the first fragment has a low median frequency of the spectrum, the second fragment has a medium frequency, the third has a high frequency, and so on cyclically. The resulting realization will fully correspond to all the criteria of the Gaussian white noise (GWN), as the original one. Neither the distribution function nor spectrum of the new realization does not differ from the original ones. But the spectrum analyzer with a sliding window equal to the length of the fragment, when shifting from left to right along the new realization, generates no longer a random but the cyclic sequence of median frequencies. Being played again in a backward direction, this realization, of course, shows the same not random sequence but in the reverse order. It has gained the detectable time-domain asymmetry (TDA).

That is why the search for the gravitational background TDA seems to us more promising than the analysis of the distribution function and spectra, because the signals of stellar mass mergers are asymmetric in the time domain.

## Signal processing method

According to our method, in order to estimate the gravitational-wave (GW) contribution to the total detectors' noise energy, it is necessary to check: is there something in this noise, which looks like the TDA of a typical GW signal? This problem must be solved only in the time domain in order to provide a linear phase response for the whole filtering system, so, preliminary spectral whitening will have to be excluded.

We have developed our filtering system built with FIR filters only since the high-order IIR filters[12] give exponentially decaying "tails" distorting the results. All processing is performed in the time domain exclusively. The weighting window realizing the nonrecursive FIR filter appropriate to solve the described problem is as follows:

$$\omega_n = -\cos\frac{2\pi n}{N-1} \cdot \sin^2\frac{\pi n}{N-1} , \qquad (1)$$

where $n = 0 \dots N-1$ is the number of the sample inside the window of length $N$ samples. Figure 1 shows the corresponding plot. The filter we applied to the original LIGO signal consists of five to eight such stages providing a linear phase response.

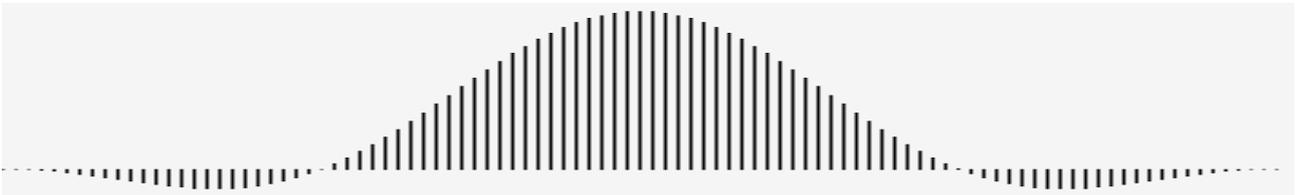

*Fig. 1. FIR filter weighting window (1) plot.*

Figure 2 shows the gravitational-wave pattern **Y** extracted from the signal **U** filtered in this way. It is frequency-limited from above but has a minimum of out-of-band interference.

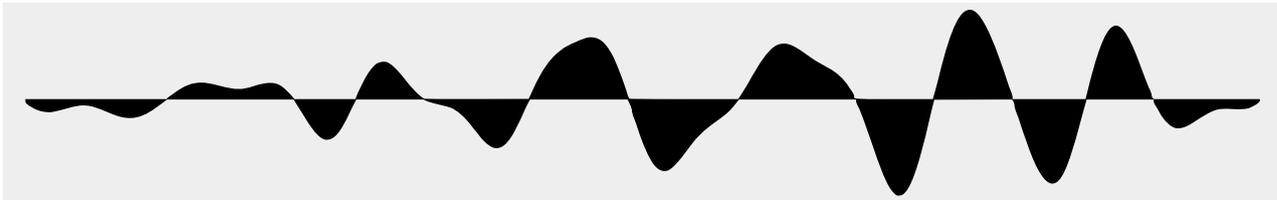

*Fig. 2. Gravitational wave pattern **Y** extracted from **U**, the original signal filtered through our filter system. The original file from the original LIGO data[13].*

Note that the original LIGO file[13] (**H-H1_GWOSC_16KHZ_R1-1185389792-32.wav**) containing the event GW170729 record[14] has the signal spectrum shown in Figure 3. Of course, at such a signal-to-noise ratio it is necessary to suppress both dominant low-frequency components and high-frequency harmonics. The inevitable non-uniformity of the frequency response in the required bandwidth would be less unfavorable for the subsequent analysis than the residual out-of-band components.

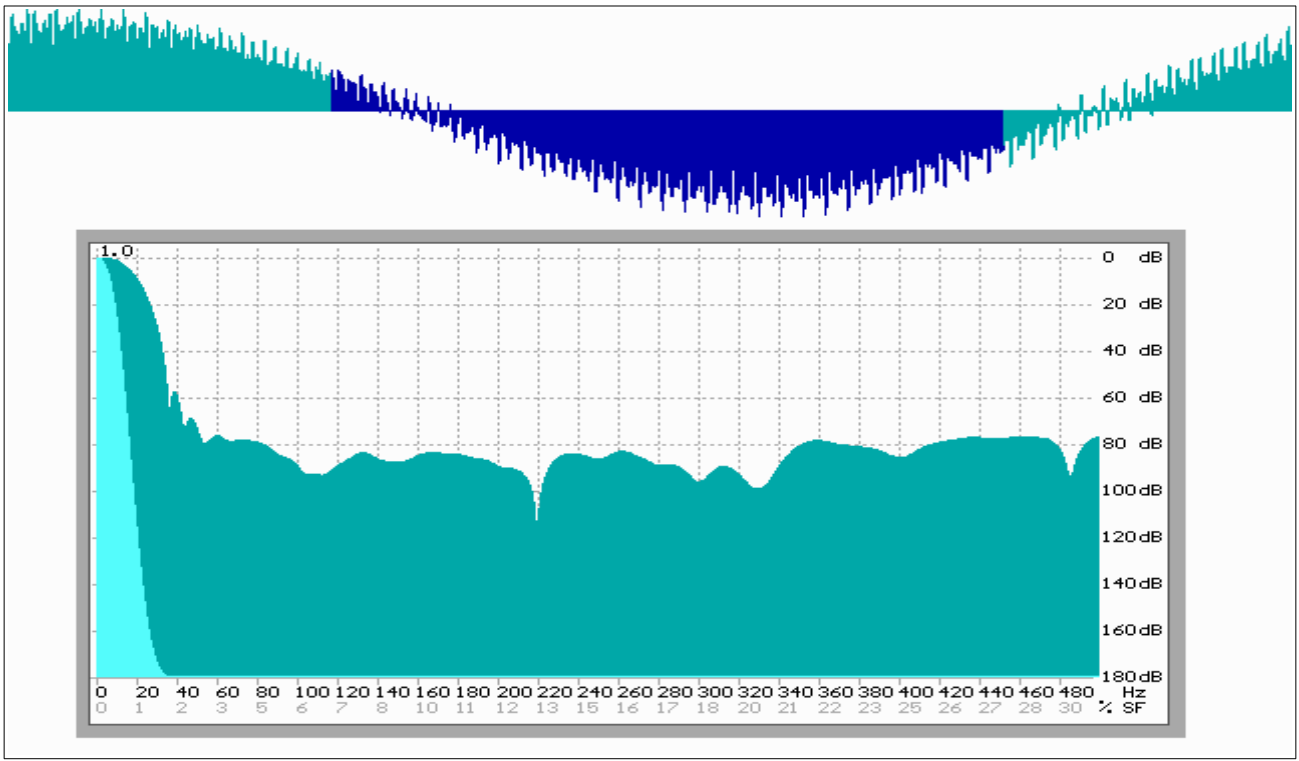

*Fig. 3. Spectral estimation of the LIGO signal. The oscillogram is at the top, the area in which the gravitational wave of the event GW170729 was recorded is highlighted in blue. The energy spectrum is at the bottom, the X-axis scale is linear, the resolution is 0.5 Hz. Hann weighting window is applied, green color is the spectrum in logarithmic scale by the ordinate, light blue color shows it in linear one.*

When filtering this, the following results were obtained (see Figure 4): −200 dB (80 + 120) in the 0...20 Hz band and at least −126 dB at frequencies above 470 Hz, which agrees well with the maximum possible suppression taking into account the residual sampling resolution. The non-uniformity of the filter frequency response in the range of 70...250 Hz (i.e. in the least noisy band) did not exceed 6 dB. The filtered signal is recorded in **H1_filt.wav** file.

We have checked the efficiency of the gravitational wave extraction and the correctness of the chosen filtering method by evaluating the operation of the detector we propose.

Detection of GW events were carried out using a specialized Pearson[15] correlation analyzer.

For the input signal from the same file **H1_filt.wav** two quantities has been calculated:

a) the square of the sliding Pearson correlation coefficient[15] $r_{XY}^2$ between current input signal fragment **X**, and **Y** reference pattern (see Figure 2);

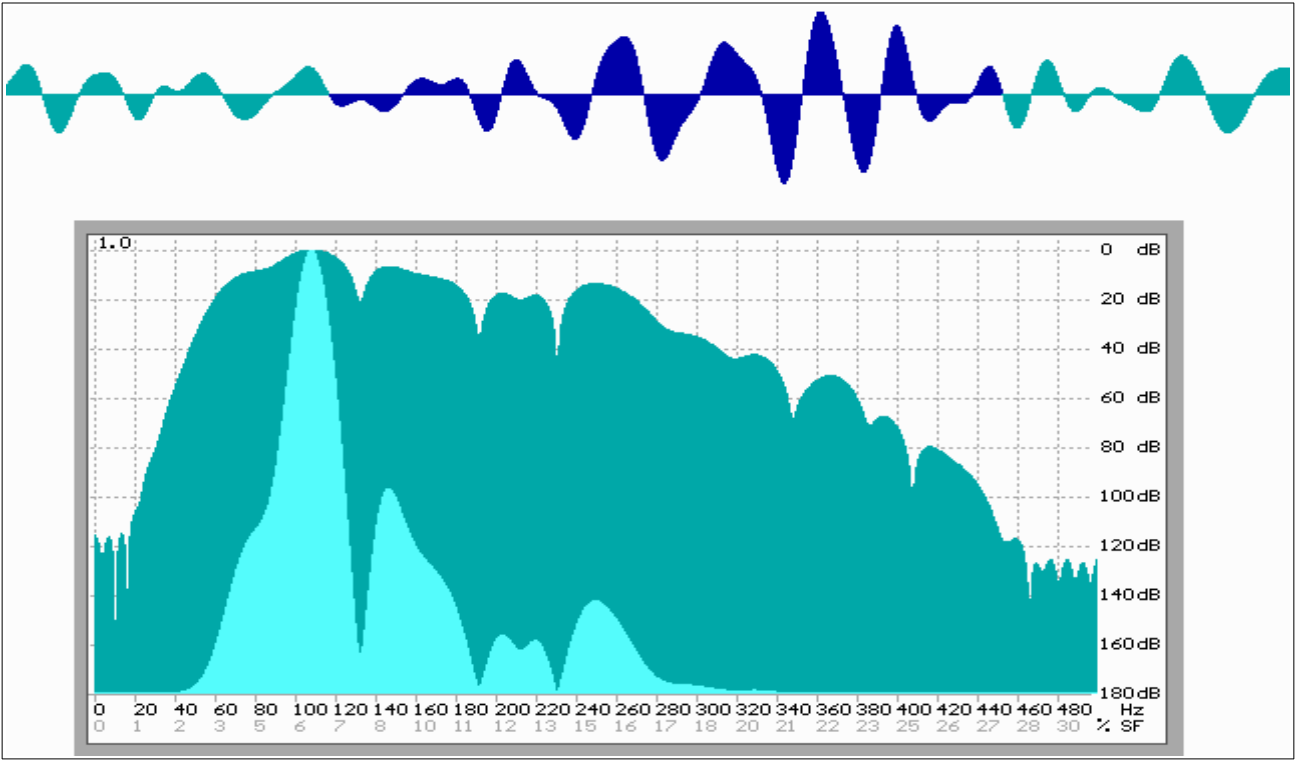

*Fig. 4. Spectral estimation of the LIGO signal after FIR filtering: the signal is confidently separated from the noise, the suppression of out-of-band components is sufficient.*

b) the square of the total sum of the current fragment **X** weighted with the cosine window[16] **W** (i.e. **X** and **W** vectors scalar product square).

The product **V** of these two quantities is forming synchronously with the input signal:

$$V_k = \left[\frac{\text{cov}_{XY}}{\sigma_X \sigma_Y}\right]^2 \times \langle \mathbf{X}, \mathbf{W} \rangle^2 = \frac{\left[\sum_{j=0}^{L-1}\left(X_j - \text{E}[\mathbf{X}]\right)\left(Y_j - \text{E}[\mathbf{Y}]\right)\right]^2}{\sum_{j=0}^{L-1}\left(X_j - \text{E}[\mathbf{X}]\right)^2 \sum_{j=0}^{L-1}\left(Y_j - \text{E}[\mathbf{Y}]\right)^2} \times \left[\sum_{j=0}^{L-1} X_j\left(1 - \cos\frac{2\pi j}{L-1}\right)\right]^2, \qquad (2)$$

where:

$L$ is the length of **W**, **X** and **Y** vectors (in number of samples); $k$ is the sliding window central sample number so that $k = i + L/2 - 1$, while $i = 0 \ldots S - L - 1$ is a current number of sliding window shifts ($S$ is the size of input signal vector **U**);

**X** is the current fragment of the input signal **U** so that $X_j = U_{i+j}, j = 0 \ldots L - 1$;

**Y** is the extracted reference pattern so that $Y_j = U_{M+j}$, where $M$ is a pattern starting sample number in the **H1_filt.wav** file; $V_k$ is the sample of the detector output signal;

σ is the standard deviation of corresponding realizations.

The $V_k$ values are always positive and will be the greater as realization of **X** correlates stronger with the **Y** pattern and the signal level of **X** is higher. As one can see from Figure 5, detector works quite well, and we used it for all the signals in this paper we work with since it has lower signal/noise ratio than a conventional convolutional detector.

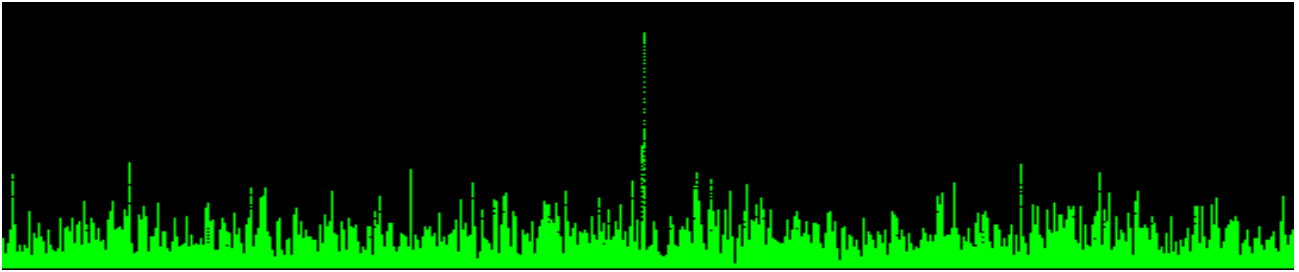

*Fig. 5. Time series of $V_k$ values calculated by formula (2) for file* **H1_filt.wav**. *The duration of the recording fragment is 24 seconds. In the middle of the graph there is a response from the "officially confirmed" wave, at least three more bursts correspond to the fragments, about which there was a discussion about the correspondence of their form to the gravitational wave pattern.*

To confirm the hypothesis about gravitational background TDA it should be experimentally recorded and, if possible, measured.

Our basic idea is that (for a sufficiently long realization) the statistically accumulated Pearson correlations of any signal symmetrical in time-domain (for example noise) with the direct and reversed reference patterns will always be equal.

Here, apparently, it is necessary to explain the idea by means of a simple model. So, let's introduce a long digitized realization of GWN, and a short fragment of signal demonstrating time-domain asymmetry: for example, a digitized sound of Euler's disk[17] falling or a chirp (LFM) radar pulse.

Now let this signal fragment slide along the realization of the noise. Calculating the Pearson coefficient for each one-sample shift, it is definitely that at any time the correlation value of the processes fluctuates between $-1.0 < r < +1.0$. The cumulative sum of all coefficients tends to be zero since the noise is not correlated with the signal. Suppose you reverse any of the signals in time, i.e. "backwards", and repeat the measurement. In that case, the result will be identical: the cumulative sum will tend to zero again, and this is expected since the noise is symmetrical in the time domain – it can be viewed from left to right or vice versa – nothing will change.

A completely different result will be obtained if the noise itself (NB!) consists of fragments of asymmetric in time signals. Suppose we have collected sounds of a very large number of simultaneously rotating and falling Euler disks. According to the central limit theorem, the total signal will be band-limited noise. Its statistical characteristics will be close to Gaussian (the closer, the more discs launched simultaneously). When calculating the correlation of this noise with a fragment of its generating signal, the following phenomenon must appear: the accumulated sum of correlations will depend on the direction of mutual displacement of the signals. And if at least one of signals is symmetric in time, there will be no asymmetry at accumulation of correlations, and the phenomenon will appear only if both (NB!) signals are asymmetric.

Let us perform processing of the real signals. To exclude the influence of signal amplitude on the result we will use only normalized correlation with the direct and reversed reference pattern. We will accumulate the difference of correlation coefficients for each pattern of the signal and present it as a graph synchronous with the work of the detector (2), i.e.:

$$R_k = R_{k-1} + r_{XY} - r_{XZ} = R_{k-1} + \frac{\text{cov}_{XY}}{\sigma_X \sigma_Y} - \frac{\text{cov}_{XZ}}{\sigma_X \sigma_Z}, \tag{3}$$

where $\mathbf{X}$ is the input signal realization, $\mathbf{Y}$ is the time-reversed reference pattern (see Figure 2), and $\mathbf{Z}$ is the direct one.

Obviously, if the signals are statistically symmetric in the time domain, the plot of $R_k$ time series will not deviate significantly either in the positive or in the negative region, i.e. accumulation will not occur. If the signal is asymmetric, the curve will shift monotonically up or down.

We also propose a method of splitting the filtered signal $\mathbf{U}$ into two components, one of which is symmetric ($\mathbf{S}$) and the other asymmetric ($\mathbf{A}$) in the time domain:

$$\begin{aligned} A_k &= r_{XY}^2 \, U_{L+i}; \\ S_k &= \left(1 - r_{XY}^2\right) U_{L+i}. \end{aligned} \tag{4}$$

Perhaps the mutual estimation of these components will allow us to determine the signal/noise ratio of the detector, if we assume the asymmetric component is true gravitational background containing bursts highly likely corresponding to unresolved GW events while the symmetric one is the sum of the hardware and induced noise.

Figure 6 shows a brief diagram depicting all the signal processing described above.

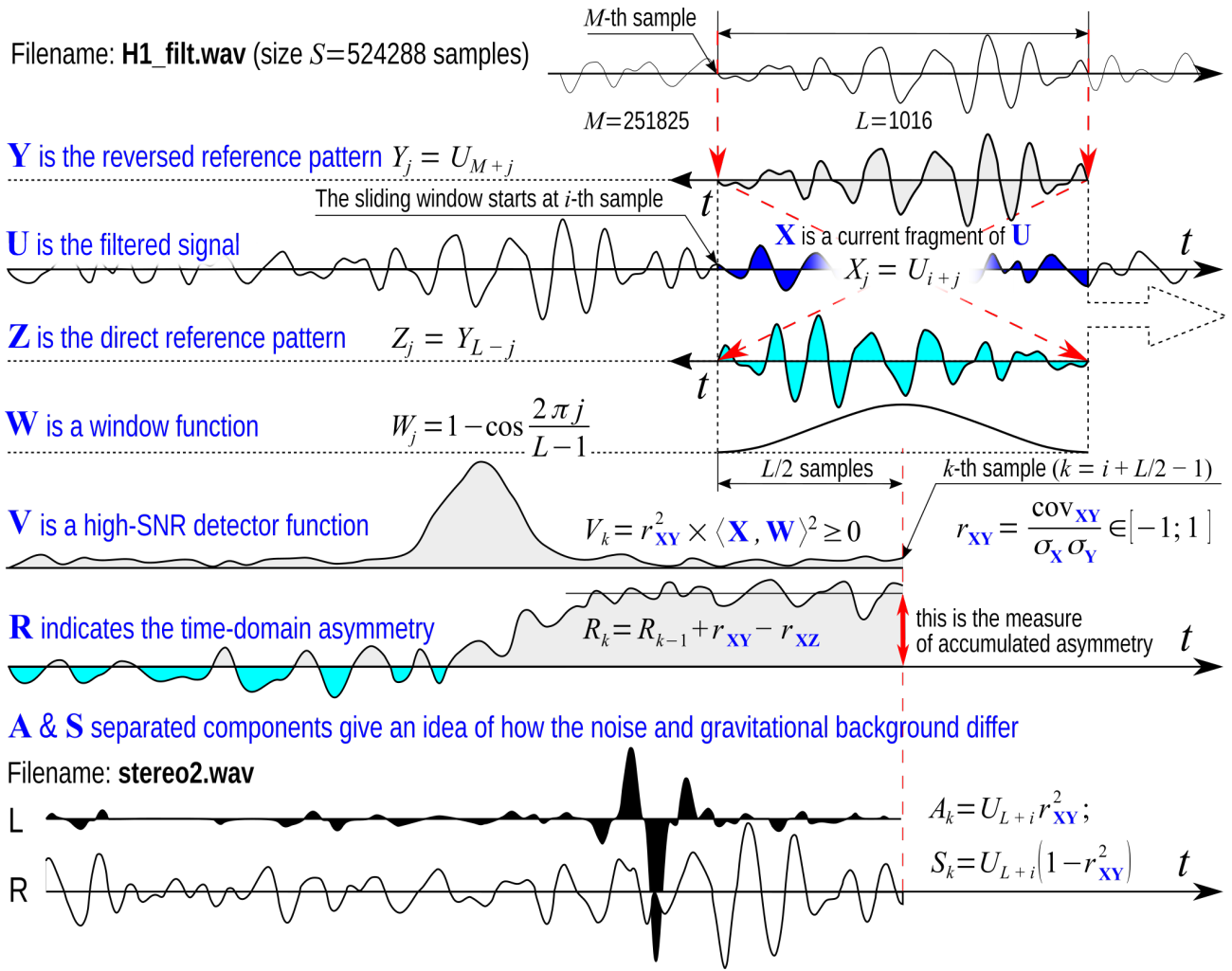

*Fig. 6. The process of GW detection and time domain asymmetry estimation brief diagram.*

**Results and discussion.**

As a zero reference, we first investigate the GWN that passes through our filtering system (file **N_filt.wav**). At the beginning and at the end of the implementation, replace the noise with a monochromatic signal for clarity and convenience in evaluating the results (see Figure 7).

Note that during the observation time (30 s) there is no accumulated asymmetry, despite periodic small deviations in both directions. The same results were observed for all other noise, unmodulated harmonic signals, and any noise/signal combinations.

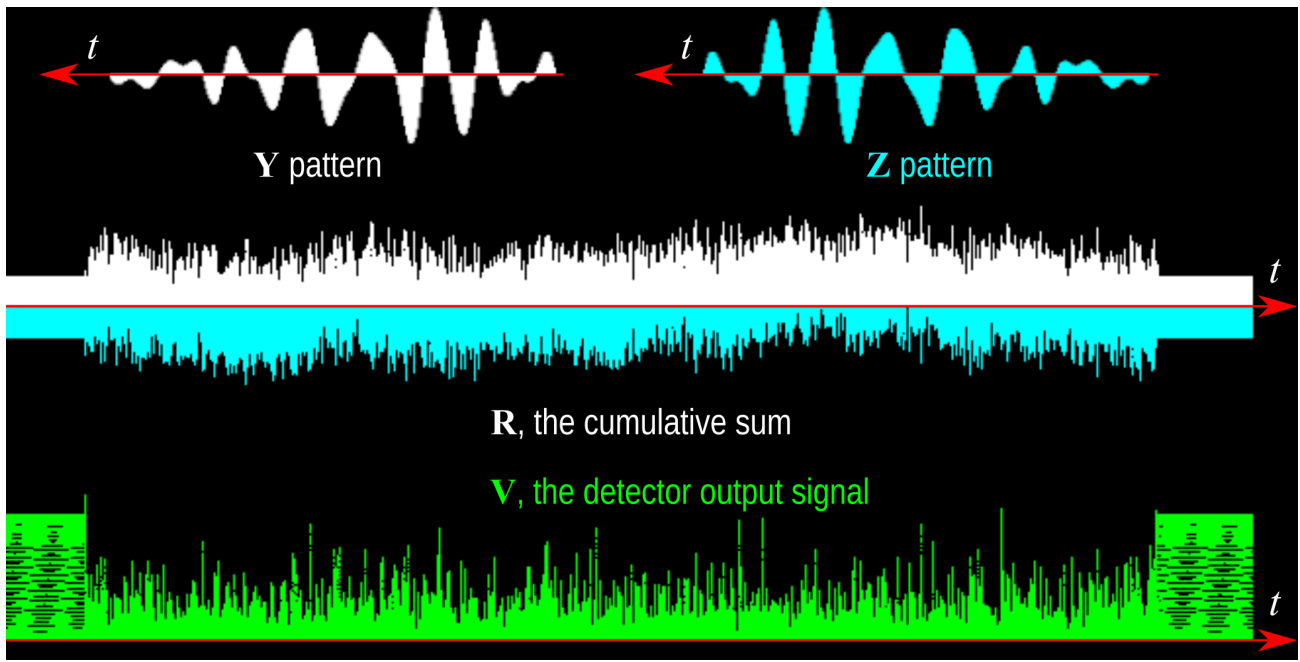

*Fig. 7. TDA estimation of the noise signal. At the top the reference patterns with which the signal is convolved: white shows the time-reversed "direct" pattern **Y**, blue indicates the normal "reversed" pattern **Z**. In the middle the cumulative sum of Pearson coefficients **R** plot is shown, i.e. the asymmetry level. The detector function **V** plot is at the bottom.*

In the search for asymmetric processes, musical works were also used, as the variety of instruments and melodies, modulation by frequency, amplitude and timbre, it would seem, should manifest itself in the asymmetry on the time scale.

The most significant fluctuations in the **R** plot are obtained in recordings of a symphony orchestra with chorus, but the accumulation of asymmetry does not occur (see Figure 8).

Completely different effects were observed in the analysis of real gravitational signals. For the

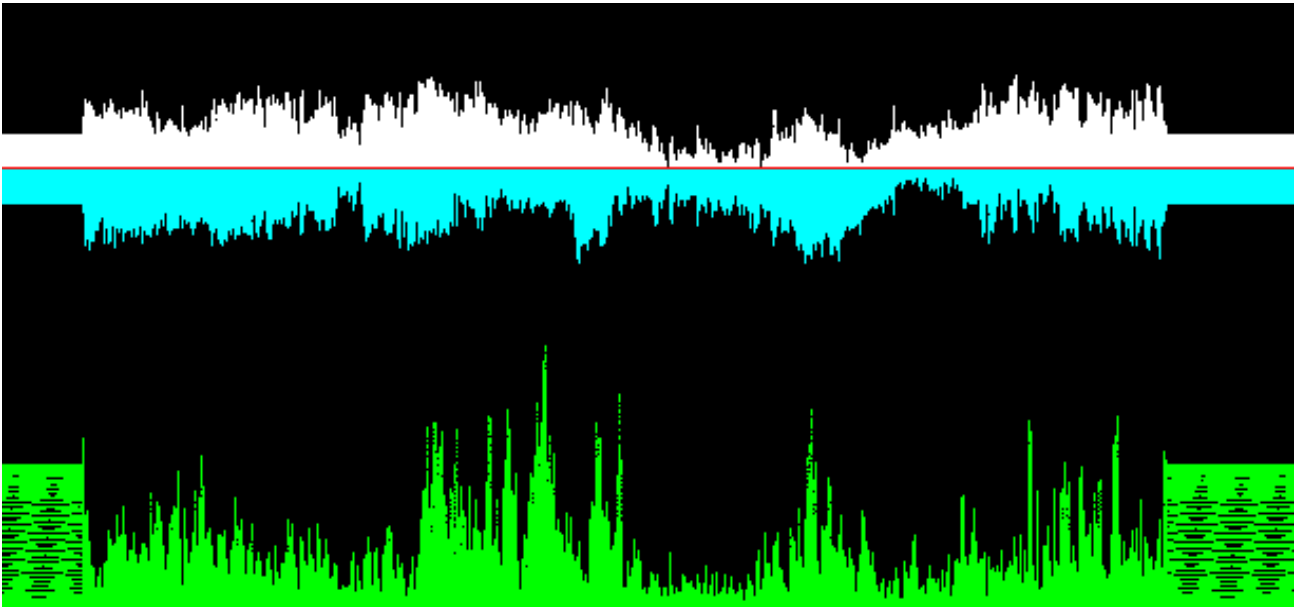

*Fig. 8. Music signal analysis. Source file: **music.wav**; filtered file: **muz_filt.wav**. Despite the fluctuations of the **R** plot, the accumulated asymmetry is equal to zero.*

Hanford Observatory, the shift of the correlation plot is confidently fixed. If we increase the observation time, the asymmetry will continue to grow monotonically (see Figure 9).

The vicinity of the most powerful gravitational wave is interesting with some anomaly lasting for units of seconds (the duration of the recording fragment is 30 seconds: the beginning and the end of the fragment are replaced by a monochromatic signal).

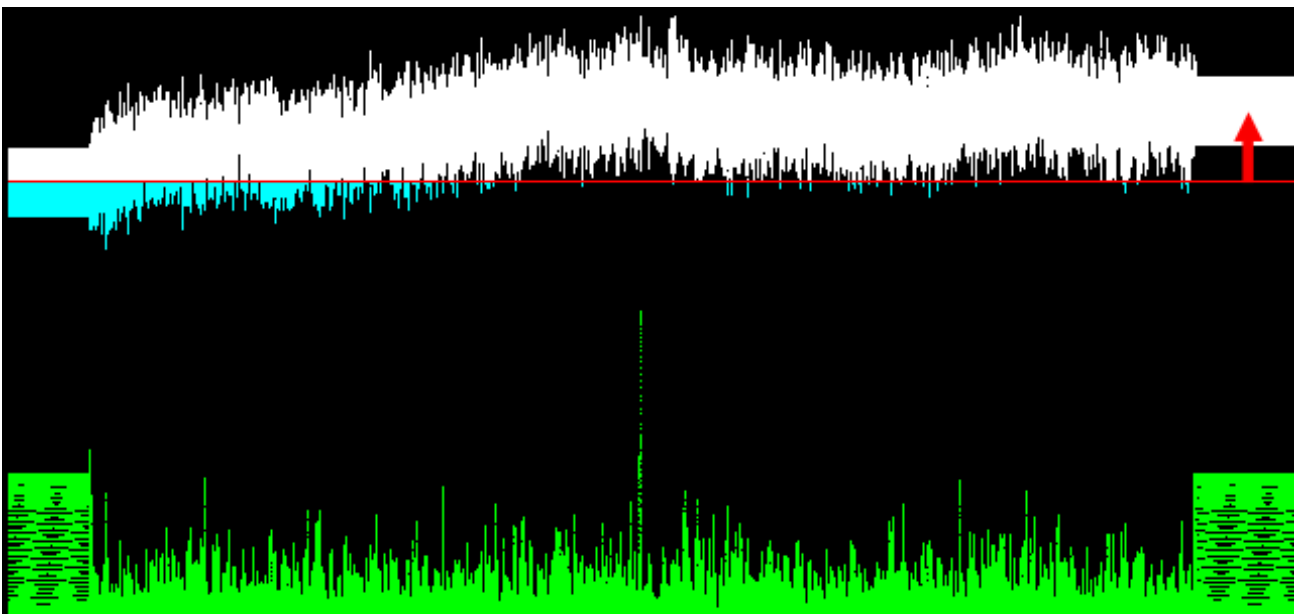

*Fig. 9. Hanford Observatory. Processing of the file **H1_filt.wav** (original **H1_GWOSC_16KHZ_R1-1185389792-32.wav**). In contrast to the noise signals, the accumulated asymmetry is not equal to zero. The red arrow shows the rise of the plot.*

For the records from the Livingstone Observatory, the asymmetry is even larger (see Figure 10). Unfortunately, detector (2) does not record any obvious bursts, which is probably due to the reference pattern, which is extracted from the Hanford recordings – the signal shapes are apparently somewhat different.

The worst, but, nevertheless, comparable results are obtained for the Virgo records (Figure 11). Unfortunately, the frequency drifting fan noise cannot be filtered out completely (the data of the compensation channels provided by the LIGO observatory[18] were not used in our work, since, unfortunately, they are not available for the event GW170729).

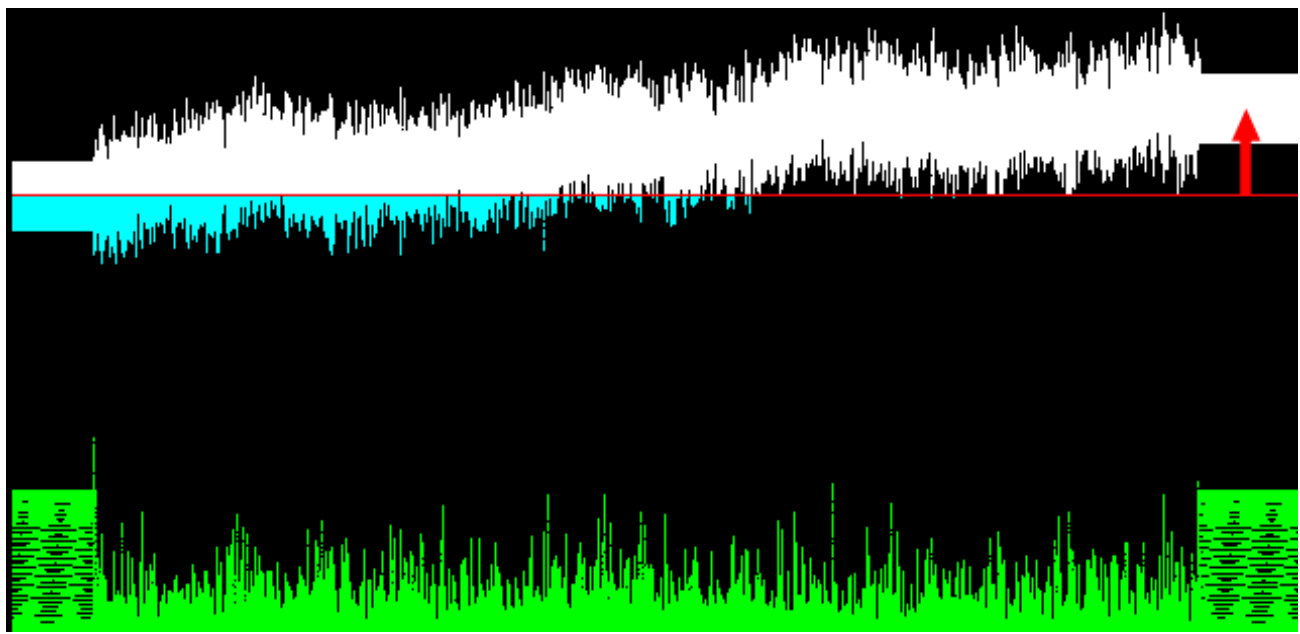

*Fig. 10. Livingstone Observatory. Processing of file* **L1_filt.wav** *(original file* **L-L1_GWOSC_16KHZ_R1-1185389792-32.wav**)*. The asymmetry is clearly observed.*

It should be noted that for the LIGO recordings, the accumulated asymmetry is unequal: all signal registrations after hardware upgrades, i.e., reduction of intrinsic noise and improvement of the signal-to-noise ratio, give a more asymmetric signal (which is to be expected).

It is impossible in principle to separate hardware, seismic, thermal, Newtonian noise from gravitational background. A basic statement of information theory states: "If signals exist simultaneously, and their spectra overlap, then complete separation of signals is impossible."

In our case, the only way to solve the problem remains: modeling signals with a known additive contribution of gravitational wave components to the structure of the GWN.

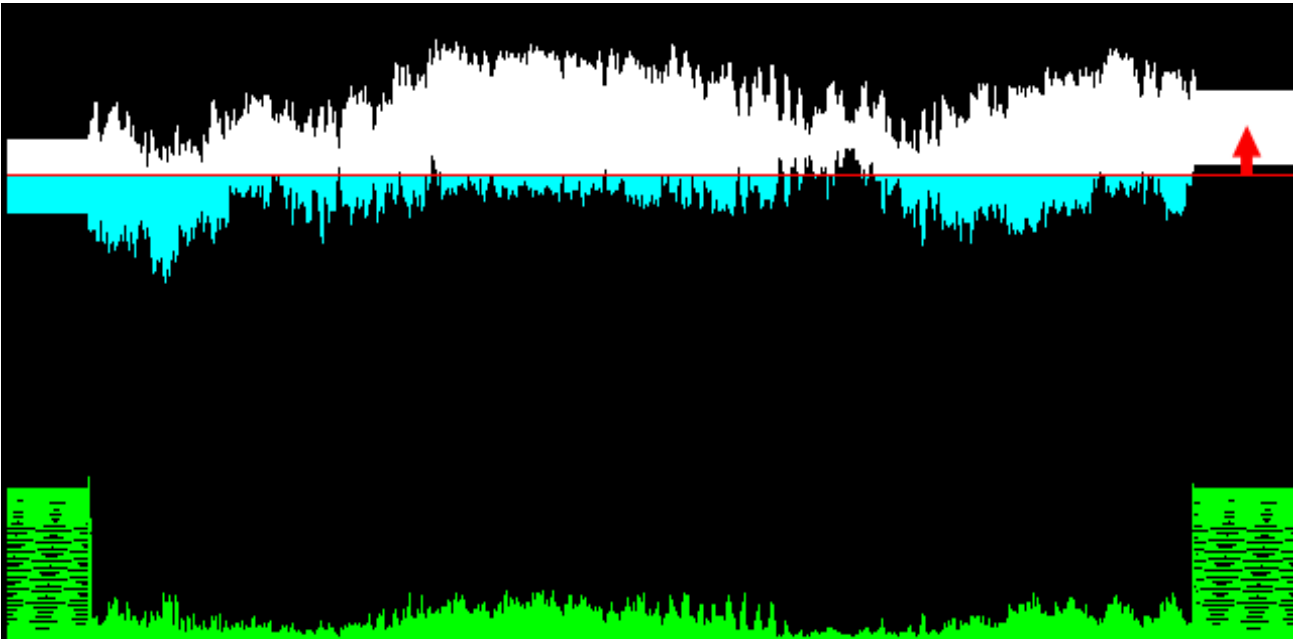

*Fig. 11. Virgo Observatory. File* **V1_filt.wav** *(original* **V-V1_GWOSC_16KHZ_R1-1185389792-32.wav***). Due to unsuppressed hardware interference, the detector does not work well. The signal asymmetry is unquestionable, but it is lower than in the previous illustrations.*

Let us make a measurement with a knownly high contribution of the "gravitational" component: let us add to the GWN fragments of the officially registered gravitational wave so that the signal to noise ratio is 3:1, (Figure 12, at such a ratio the signal against the noise will already be observed visually).

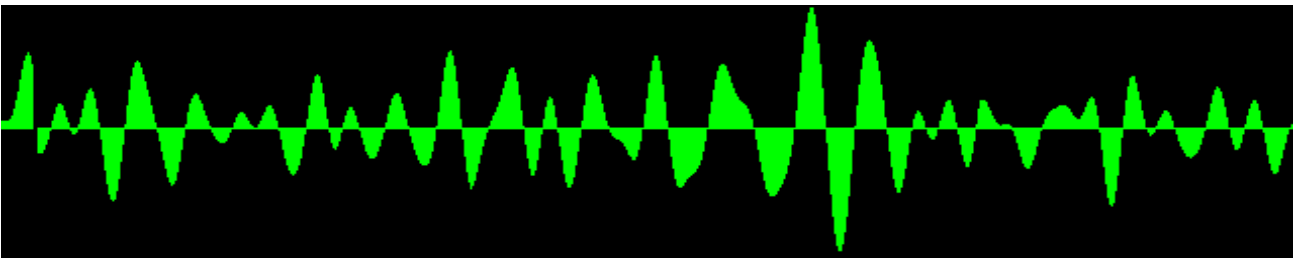

*Fig. 12. Additive mixture of the components of gravitational waves and GWN in the ratio 3:1. As expected, the characteristic patterns are visually detectable.*

Let us estimate the accumulation of asymmetry and monitor the detector performance (Figures 13 and 14). We will assume that the measuring instrument is pre-tested, and it is possible to select such a signal-to-noise ratio, which will allow to reach the values of the curve rise as in the cases of gravitational detector signal recordings.

Here we can add that a partial separation of the noise components is possible in the time domain, i.e. the signal is separated into two channels using equation (4): an asymmetric component (in the left channel of the **stereo2.wav** file), which corresponds to a possible gravitational background, and a symmetric one (in the right channel), probably containing only noises (see Figure 15). The **stereo2.wav** file, when listened to, gives an idea of how the pure noise and gravitational background components differ (high-amplitude pulses in the left channel at the 15th second correspond to the event GW170729). We also are pleased to present the working Windows version of the analyzer program that we developed, with the help of which we obtained the results published here. A program (together with the brief instruction on how to use) is inside the **archive file**. The file size is 23 M.

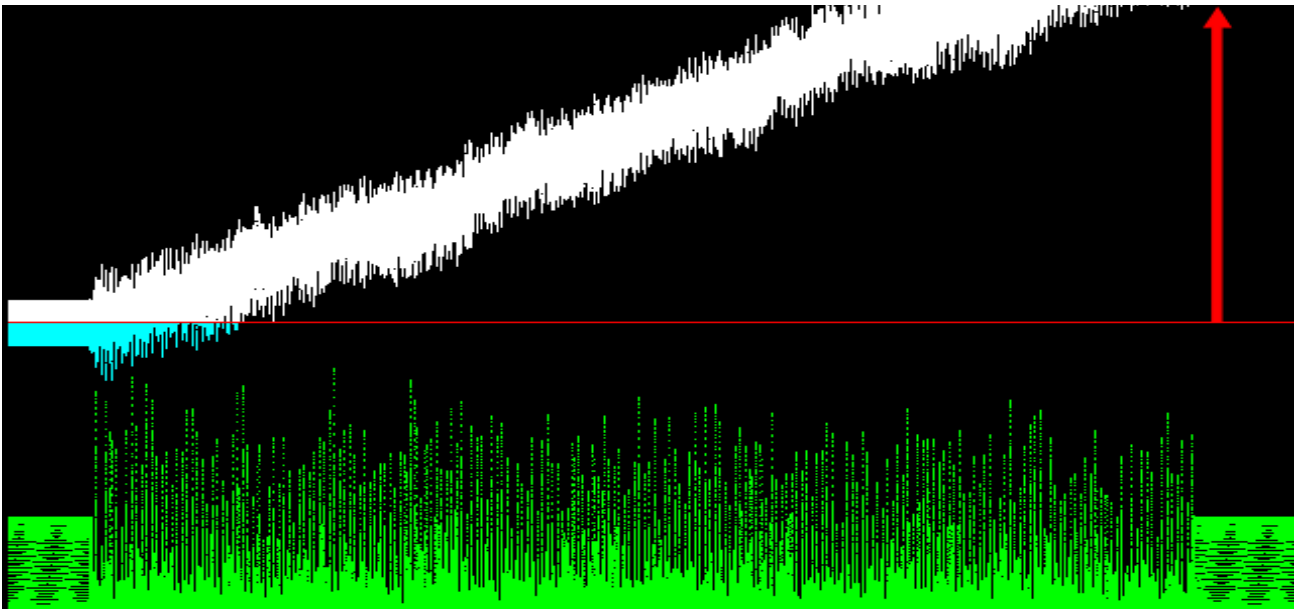

*Fig. 13. Analyzer performance with a significant contribution of the "gravitational" component. The signal-to-noise ratio is 3:1. Multiple **V** spikes are detected. The asymmetry is large, the accumulated difference of Pearson coefficients grows rapidly.*

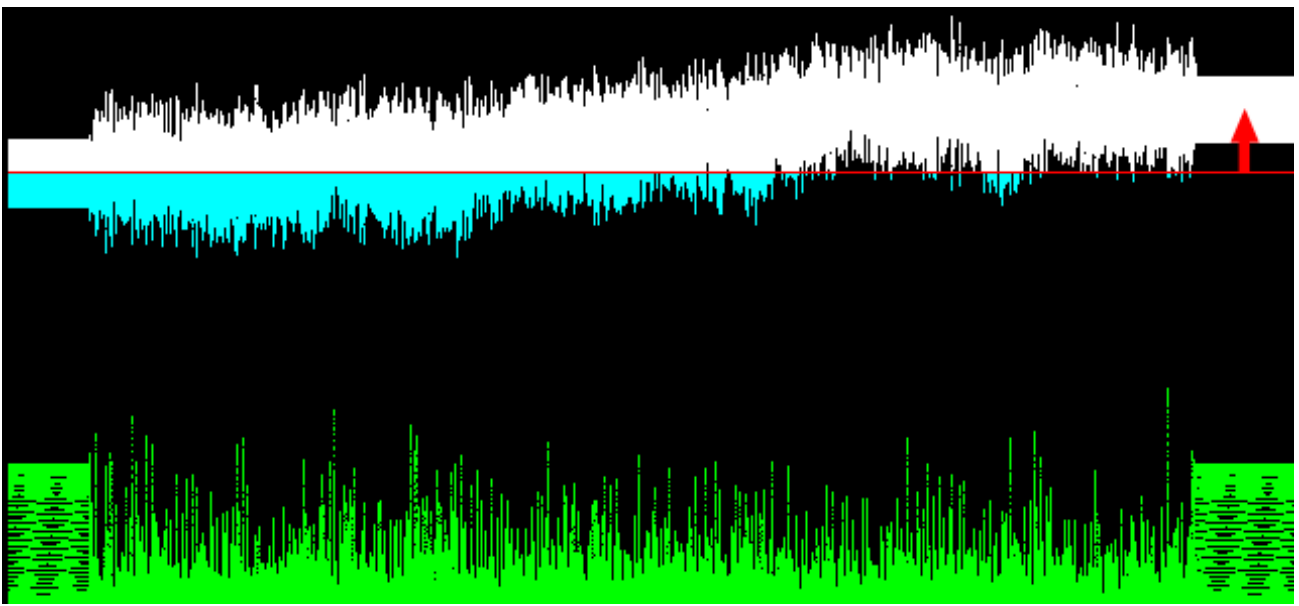

*Fig. 14. The signal-to-noise ratio is 1:2. Complete correspondence to the average value of the asymmetry for the records of gravitational detectors. The estimate is, of course, preliminary.*

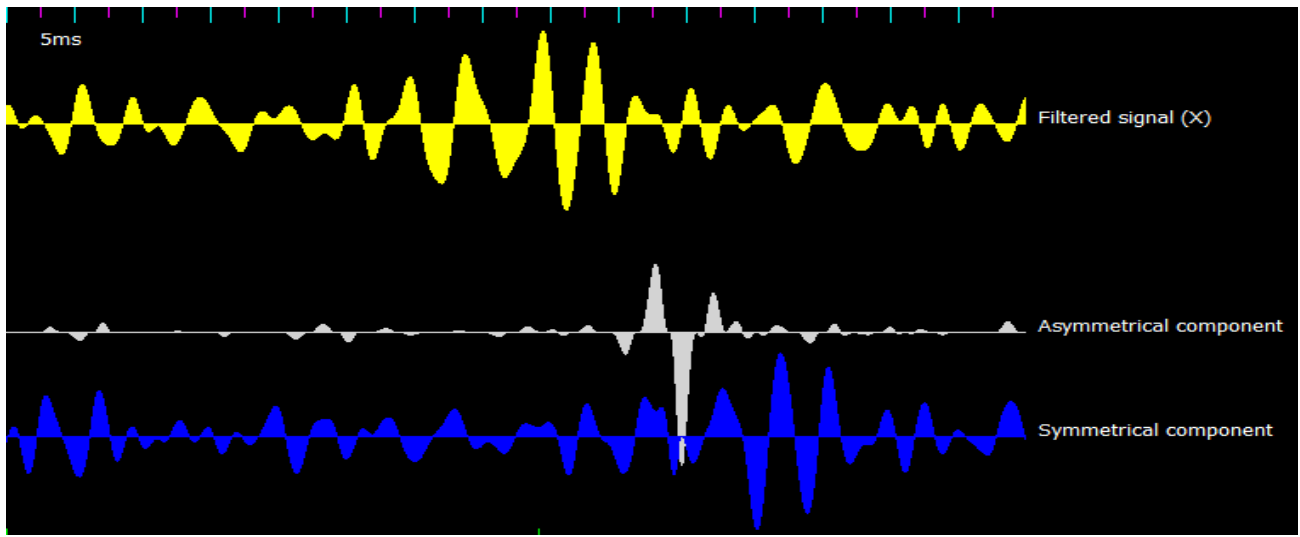

*Fig. 15. The signal of event GW170729 after the filter (1) and its asymmetric and symmetric (in time) components. The time delay in the appearance of symmetric signals can be clearly seen. The reason for this phenomenon is not yet clear to the authors, but we can assume that the appearance of delayed signals due to the interaction of the gravitational wave with the mass of the Earth.*

The dataset containing all files mentioned in this paper is presented here[19]

## Conclusions

The results of indirect measurements allow us to estimate the gravitational background level at the output of the detectors is −6 dB approximately, and this, in fact, is an unexpectedly[10] large value. So, the essential part of the permanent noise is due to the same processes as the rarely observed high-amplitude gravitational waves, namely the merging of massive astronomical objects. The very presence of asymmetry of the signal received by detectors means that the whole space is filled with gravitational noise of the sub-kilohertz band. And, consequently, any optical and radio-frequency signals always and everywhere propagate in the gravitational-noise field.

Then the input signal of any radio telescope directed to any point of space will include the ripple caused by gravitational background[20,21] and manifested as a differential modulation by azimuth and angle of location in the observer's picture plane (i.e. change of scattering ellipse and/or polarization).

Accordingly, we predict that at the output of the quadrature polarimeter having measurement time less than 1 ms and bandwidth limitation at frequencies below 70 Hz there will be an additive mixture of its intrinsic noise and gravitational background component, which can now be estimated (although it is still impossible to completely separate these noises). The contribution of gravitational background may be small, but with increasing of realization, i.e. registration time, the measurement accuracy also will increase. Then maps of the gravitational background noise of the sky can be obtained with the available radio telescopes. In addition, the relic radiation should also have traces of propagation in such a gravitational background noise field.